\documentclass[%
reprint,
 amsmath,amssymb,
 aps,
]{revtex4-2}

\usepackage{graphicx}
\usepackage{dcolumn}
\usepackage{bm}
\usepackage[colorlinks=true,linkcolor=black]{hyperref}

\usepackage{derivative}
\usepackage{enumitem}
\usepackage{array}
\usepackage{amsthm}
 
\newcolumntype{L}[1]{>{\raggedright\arraybackslash}p{#1}}

\newtheorem{lem}{Lemma}

\usepackage{xr}
\makeatletter
\newcommand*{\addFileDependency}[1]{
\typeout{(#1)}
\@addtofilelist{#1}
\IfFileExists{#1}{}{\typeout{No file #1.}}
}\makeatother
\newcommand*{\myexternaldocument}[1]{%
\externaldocument{#1}%
\addFileDependency{#1.tex}%
\addFileDependency{#1.aux}%
}
\myexternaldocument{si}

\begin{document}
\title{Hebbian Physics Networks: A Self-Organizing Computational Architecture Based on Local Physical Laws}
\author{Gunjan Auti}
\email{gunjanauti@thml.t.u-tokyo.ac.jp}
\affiliation{Department of Mechanical Engineering, The University of Tokyo, Japan}

\author{Gouhei Tanaka}
\affiliation{Graduate School of Engineering, Nagoya Institute of Technology, Japan}
\affiliation{International Research Center for Neurointelligence, The University of Tokyo, Japan}

\author{Hirofumi Daiguji}
\affiliation{Department of Mechanical Engineering, The University of Tokyo, Japan}

\begin{abstract}
Physical transport processes organize through local interactions that redistribute imbalance while preserving conservation. Classical solvers enforce this organization by applying fixed discrete operators on rigid grids. We introduce the Hebbian Physics Network (HPN), a computational framework that replaces this rigid scaffolding with a plastic transport geometry. An HPN is a coupled dynamical system of physical states on nodes and constitutive weights on edges in a graph. Residuals, local violations of continuity, momentum balance, or energy conservation, act as thermodynamic forces that drive the joint evolution of both the state and the operator (i.e., the adaptive weights). The weights adapt through a residual-modulated anti-Hebbian rule, which we prove constitutes a strictly local gradient descent on the residual energy. This mechanism ensures thermodynamic consistency: near equilibrium, the realized transport operator converges to a symmetric, positive-definite form, reproducing Onsager’s reciprocal relations without explicit enforcement. Far from equilibrium, the system self-organizes into transport topologies within a thermodynamically admissible class that redistribute imbalance through adaptive local geometry. Unlike optimization-based approaches that impose physics through global loss functions, HPNs embed conservation intrinsically: transport is restored locally by the evolving operator itself, without a global Poisson solve or backpropagated objective. We demonstrate the framework on scalar diffusion and incompressible lid-driven cavity flow, showing that physically consistent transport geometries and flow structures emerge from random initial conditions solely through residual-driven local adaptation. HPNs thus reframe computation not as the solution of a fixed equation, but as a thermodynamic relaxation process in which the constitutive geometry and physical state co-evolve.
\end{abstract}


\maketitle
\section{Introduction}
The central abstraction of classical computational physics is the separation of state and structure. Classical continuum modeling assumes that the state $U(x,t)$ evolves under a prescribed local evolution operator, 
\begin{equation}
    \frac{\partial U}{\partial t} = F(U, \nabla U, \nabla^2U,...; x, t, \theta),
    \label{eq:delU_delt}
\end{equation} 
where $F$ (including constitutive closures $\theta$) is fixed \textit{a priori} and the computation advances the state $U$ forward in an externally defined time variable. In a conventional simulation, this assumption is realized discretely. The physical state, such as velocity, temperature, or concentration, evolves in time, while the geometric structure governing its transport is fixed in advance. Whether defined by a finite difference stencil, a finite element stiffness matrix, or a connection graph, the discretized constitutive operator is treated as a rigid scaffolding. Conservation is enforced by repeatedly applying this fixed operator under a globally synchronized update, forcing the system to conform to a prescribed geometric constraint \cite{patankar2018numerical, ferziger2019computational, leveque2002finite}.

Classical methods remain highly successful far-from-equilibrium, but they achieve this by embedding non-equilibrium effects within fixed or parametrized operators. In turbulent flows, effective momentum transport is represented through state-dependent eddy viscosities defined within prescribed closure frameworks \cite{boussinesq1877essai, pope2001turbulent}. In reacting systems, transport and reaction pathways reorganize through state-dependent rates, while the underlying interaction structure remains externally specified \cite{de2013non}. Similarly, in electrochemical and electrostatic systems, conductivities and screening lengths vary with local charge and potential, yet are incorporated through constitutive relations rather than through adaptation of the transport geometry itself \cite{bazant2009towards}. In all such cases, non-equilibrium behavior is accommodated by enriching or parameterizing the function $F$, while retaining the deeper assumption that the transport structure itself is fixed and that the system’s dynamics remain closed in the state variables alone.

This separation between evolving state and fixed structure also underlies a range of adaptive and local computational approaches. Energy-based and relaxation-based models interpret computation as descent on an internal Lyapunov functional, but typically operate on a fixed interaction topology \cite{Hopfield1982, LeCun2006, Friston2010}. Pseudo-time and artificial-compressibility methods introduce internal relaxation clocks to enforce constraints locally, yet rely on prescribed operators \cite{chorin1997numerical, jameson1991time}. Lattice-based \cite{frisch1986lattice, mcnamara2019use} and cellular-automaton \cite{wolfram1986theory} approaches generate macroscopic transport from local update rules, though these rules are fixed \textit{a priori}. Adaptive discretization strategies, such as mesh refinement or coefficient tuning, modify numerical resolution or parameters while leaving the constitutive topology unchanged. Across these approaches, adaptation occurs within a \textit{fixed} transport structure.

From the perspective of non-equilibrium thermodynamics, however, this fixed-structure assumption is not fundamental. Onsager’s reciprocal theory identifies the symmetric, positive operator linking thermodynamic forces to fluxes as the local geometric structure governing dissipation \cite{Onsager1931, callen1993thermodynamics}. Prigogine’s extension to far-from-equilibrium systems emphasizes that conservation is maintained through continual local adjustment of these force–flux couplings, rather than through global coordination \cite{Prigogine1977, Prigogine1980}. In this view, the constitutive operator is not a static background object but a dynamic field that co-evolves with the state variables as thermodynamic stress is redistributed.

\subsection*{Description of HPN}

Building on this thermodynamic picture, we introduce the Hebbian Physics Network (HPN), a local relaxation solver based on the idea that physical evolution proceeds through the local restoration of broken conservation laws. Any external disturbance---such as forcing, boundary motion, or initial heterogeneity---locally disrupts the balance of conserved quantities, even though global conservation remains intact. Non-equilibrium systems respond to such imbalances by generating fluxes from their immediate surroundings, redistributing mass, momentum, or charge until local balance is restored, a mechanism underlying relaxation and pattern formation in driven systems \cite{cross1993pattern}. HPN departs from this paradigm by abandoning the assumption that the dynamics are governed by a fixed evolution operator: here, the effective evolution law is not prescribed, but constructed dynamically through local adaptation of the transport structure itself.

Crucially, the specific dynamical formulation by which this redistribution occurs is not assumed \textit{a priori}. Instead, the HPN treats the mechanism of transport itself as an adaptive degree of freedom: transport is represented as a network of local pathways whose strengths reorganize in response to imbalance. When a local violation of conservation arises, nearby pathways adjust to facilitate redistribution, and the state relaxes along these evolving connections. In this formulation, physically consistent behavior emerges not by explicitly time-integrating a fixed evolution equation of the form $\partial_tx = f(x)$ [Eq.~\eqref{eq:delU_delt}], but by allowing the transport structure itself to adapt until local conservation is restored. The effective dynamics of the system are therefore encoded in the evolving transport geometry, rather than imposed through a prescribed governing equation.

\begin{figure}
\centering
\includegraphics[width=\linewidth]{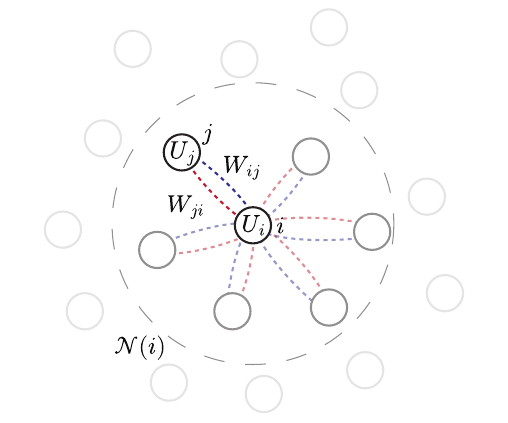}
\caption{\textbf{Schematic of the graph $\mathcal{G}$}: each node $i$ interacts with its neighbors $j\in\mathcal N(i)$ through directed couplings $W_{ij}$; reciprocal edges $j\to i$ carry distinct couplings $W_{ji}$. $U_i$ and $U_j$ are the dynamic state variables at nodes $i$ and $j$, respectively.}
\label{fig:graph}
\end{figure}

The network is defined on a discrete graph whose nodes store dynamic state variables ($U$) and whose directed edges represent the pathways along which flux can flow. Each directed edge $j \to i$ carries a coupling weight $W_{ij}$ that determines how the state at node $j$ contributes to the flux entering node $i$ (see Fig. \ref{fig:graph}). These weights collectively form the local geometric structure of the graph: they specify which transport pathways are available and how strongly they interact.

To drive the adaptation of this geometry, we identify the local imbalance measured internally by the network. For each node $i$, we compute a residual using discrete Gauss Divergence theorem:

\begin{align}
R_i &= \sum_{j} J_{ij} - \sum_j J_{ji}, \nonumber \\
 \quad J_{ij}&= W_{ij}~S_{ij} ,
\label{eq:residual_def}
\end{align}
where, $S_{ij}$ is the local signal from node $j$ to node $i$. The choice of $S_{ij}$ is dictated by the constitutive relation of the target physics. Residual  $R_i$ is the physical accumulation of the conserved quantity (e.g., mass or momentum) at node $i$ as shown in Eq~\eqref{eq:residual_def}. This accumulation creates a local stress that acts upon the incident edges. It quantifies a local deviation whose relaxation requires modifying the force--flux mapping. Therefore, serving as the natural thermodynamic signal driving the dynamics of the constitutive operator. The nodal state 
$U_i$ represents an intensive density (detailed definition in Sec.~\ref{sec:state_def}) stored in a node of finite capacity $\Omega_i$. Accordingly, the state evolves by relaxing this accumulation at a rate inversely proportional to the local capacity. The state update is therefore defined as:
\begin{align}
\Delta U_i = \frac{\eta_u}{\Omega_i} \left( \sum_{j \in \mathcal{N}(i)} (J_{ij} - J_{ji})\right) = \frac{\eta_u}{\Omega_i} R_i,
\end{align}
where, $\eta_u$ is the kinetic relaxation rate defined by a systemic stability constraint such as a Courant–Friedrichs–Lewy (CFL)-like limit. 

The system adheres to a principle of constitutive relaxation: when a conserved quantity accumulates at a node (i.e. residual $R_i$), the incident pathways `yield' by augmenting their effective conductance to alleviate local stress, whereas, pathways opposing this relaxation are suppressed. This reshaping of the transport geometry ensures that the network converges toward a configuration where local conservation is satisfied intrinsically. This mechanism is formalized as a residual-modulated anti-Hebbian rule in form \cite{foldiak1990forming, fremaux2016neuromodulated, isomura2018error}:
\begin{equation}
\Delta W_{ij} = -\eta_{w}
f(R_i) S_{ij} - \lambda W_{ij},
\label{eq:hebbian_rule}
\end{equation}
where the update is governed by: (i) a signal basis $S_{ij}$, which defines the local transport context (pre-synaptic activity), (ii) a modulatory residual $f(R_i)$ representing the local violation of balance, and (iii) a systemic structural mobility $\eta_{w}$. 

Importantly, this defines a coupled dynamical process in which the transport geometry evolves continuously in response to local imbalance, while global conservation is satisfied algebraically at all times (i.e., $\sum_i R_i = 0$). The directed couplings $W_{ij}$ adapt their transmission capacity to relieve local accumulation, reshaping the effective transport pathways wherever imbalance persists. This process does not presuppose convergence to equilibrium: if external driving, boundary forcing, or sustained heterogeneity continually generate local stress, the network will continue to reorganize its transport structure. As a result, the HPN admits equilibria, non-equilibrium steady states, and time-dependent regimes within a single unified formulation.

Crucially, the resulting evolution is not an arbitrary search in configuration space. Because imbalance is measured through conservation residuals and relaxation occurs through strictly local updates, each step of the dynamics redistributes conserved quantities while adjusting the constitutive geometry that mediates that redistribution. In Section~\ref{sec:theory}, we show that these updates correspond to a local descent on a residual energy functional and recover reciprocal transport structure in the near-equilibrium limit. The trajectory generated by the HPN is therefore constrained by conservation and local dissipation, rather than being imposed by an externally prescribed time-marching scheme. In this sense, the object computed by an HPN is the relaxation path defined by these constraints, not merely a terminal steady state.

The remainder of this paper develops the formal structure of the HPN and demonstrates its operation on canonical transport problems. Section~\ref{sec:theory} defines the local fluxes and establishes that the Hebbian update constitutes a local gradient descent on the residual energy. Section~\ref{sec:diffusion} examines scalar transport, illustrating how a consistent constitutive operator emerges solely through residual-driven feedback. Section~\ref{sec:cavity} extends the framework to incompressible lid-driven cavity flow, where pressure enforces the divergence constraint while the momentum-exchange weights adapt to maintain the correct local structure. Finally, in Section~\ref{sec:discussion} we discuss how physically consistent transport can arise from local constraint enforcement without a fixed operator. The HPN does not impose a constitutive geometry in advance; it learns one by continually adjusting the force--flux mapping so that imbalance is eliminated \textit{in situ}.

\section{Thermodynamic Learning via Local Residuals}
\label{sec:theory}

Having established the qualitative framework of a \textit{plastic} transport geometry, we now formalize the mechanics of adaptation. In this section, we define the Hebbian Physics Network (HPN) as a coupled dynamical system $(U, W)$ and analyze the local relaxation principles governing its evolution. We show that, for a broad class of residual-driven update rules, the adaptation of the transport weights constitutes a strictly local gradient descent on a well-defined residual energy, ensuring thermodynamic admissibility of the constitutive dynamics.

Crucially, the results in this section characterize the \textit{structural} properties of the HPN update rules rather than guaranteeing convergence to a particular physical solution in all settings. Under near-equilibrium conditions and appropriate separation between state relaxation and structural adaptation, the learned transport operator approaches a symmetric, linear response form consistent with Onsager’s reciprocal relations. Away from this limit, the same local adaptation mechanism governs the co-evolution of state and transport geometry without presupposing global equilibrium or a fixed constitutive law.

\subsection{Formal definitions of State $U$ and Signal $S$}
\label{sec:state_def}
To map a physical system onto the HPN substrate, we must identify the fundamental degrees of freedom that characterize the local state and the nature of the interactions between them. The state of node $i$, denoted by $U_i$, represents the density of a quantity subject to a global conservation law. In the context of dynamical systems, $U_i$ is the primary degree of freedom for which a local conservation law, such as the conservation of mass, momentum, or energy, is defined. For example, in a dissipative process such as diffusion, $U_i$ represents the local concentration $c_i$. In fluid dynamics, $U_i$ represents the momentum density $\rho \mathbf{u}_i$, where $\rho$ is the fluid density and $\mathbf{u}_i$ is the velocity vector. 

Crucially, in the HPN framework, global conservation is a structural invariant of the graph architecture. Because the network only redistributes the state among nodes via directed edges, the total sum $\sum_i U_i$ remains algebraically constant throughout the evolution. What the HPN focuses on is the local breaking of conservation, quantified by the residual $R_i$, and the subsequent restoration of local symmetry through autonomous relaxation.

The interaction between nodes is mediated by a coupling signal $S_{ij}$, representing the physical information transmitted from node $j$ to node $i$ across a directed edge. In the HPN framework, we adopt a state-centered signal,
\begin{equation}
    S_{ij} = U_j,
\end{equation}
and encode the constitutive physics entirely in the adaptive operator $W_{ij}$. The operator acts on the signal to generate a directed flux described by Eq.~(\ref{eq:residual_def}), which governs the redistribution of the conserved quantity across the graph.

With directed edges $i \to j$ and $j \to i$ treated as distinct pathways, conservation is enforced through the nodal residual,
\begin{equation}
    R_i = \sum_i \left(W_{ij}S_{ij} - S_{ji}W_{ji}\right),
\end{equation}
which represents the net accumulation of the conserved quantity at node $i$. This form corresponds to a discrete continuity equation on a directed graph: the first term accounts for incoming flux, while the second represents outgoing transport proportional to the local state. Diffusive and advective behavior are not distinguished by the signal, but by the symmetry properties of the learned operator. In the reciprocal limit $W_{ij} \approx W_{ji}$, the residual reduces to a graph Laplacian and recovers diffusive transport; deviations from reciprocity encode advective bias and circulation.

Crucially, the HPN does not learn isolated physical coefficients. Instead, $W_{ij}$  evolves into a composite transport operator that captures the effective geometry through which imbalance is redistributed. In complex systems such as incompressible flow, the residual $R_i$ aggregates contributions from multiple physical mechanisms (e.g., advection, dissipation, and constraints). The adaptive operator reorganizes to resolve this total thermodynamic stress without requiring separate, explicitly prescribed operators for each term.

\subsection{Problem statement and notation}
\label{sec:problem}
We consider a directed graph $\mathcal{G}=(\mathcal{V},\mathcal{E})$ where each node $i \in \mathcal{V}$ carries the intensive state $U_i \in \mathbb{R}^d$ and each edge $(i,j) \in \mathcal{E}$ carries a coupling operator $W_{ij} \in \mathbb{R}^{d\times d}$ representing its geometric conductance. The edge set $\mathcal{E}$ defines the fixed connectivity of the system and is constructed from the physical embedding of the nodes by a local neighborhood criterion (e.g., a radius-based or 
$k$-nearest-neighbor rule), which enforces locality but does not prescribe a precise metric within the neighborhood. To bridge the topological structure with physical space, we introduce a diagonal nodal metric matrix $\mathbf{G} = \text{diag}(\Omega_1, \dots, \Omega_N)$. Here, $\Omega_i > 0$ denotes the local material capacity or nodal volume, ensuring the total conserved quantity of the substrate is defined by the weighted sum $\mathbf{1}^\top \mathbf{G} \mathbf{U}$.

The neighborhood structure is determined directly from the known spatial coordinates of the nodes. Specifically, two nodes \(i\) and \(j\) are connected by a directed edge if their coordinates satisfies a Chebyshev distance criterion,
\[
\lVert \mathbf{x}_i - \mathbf{x}_j \rVert_{\infty} \le k,
\]
where \(k\) is a prescribed interaction radius. This construction yields a fixed, local interaction graph that enforces spatial locality while remaining agnostic to the precise transport metric within the neighborhood; all constitutive transport detail is subsequently encoded through the adaptive evolution of the edge operators \(W_{ij}\).

The local dynamics are driven by the interaction between the operator and the signal $S_{ij}$. This interaction generates a directed flux $J_{ij}=W_{ij}S_{ij}$, the accumulation of which defines the nodal residual as defined by Eq.~(\ref{eq:residual_def}). Physically, $R_i$ represents a local thermodynamic stress indicating a departure from equilibrium. Because the residual is constructed purely from internal flux redistribution, global conservation is a structural invariant of the architecture, satisfying $\sum_i R_i = 0$ regardless of the operator configuration.

To characterize the collective relaxation of the substrate, we define the global state $\mathbf{U} \in \mathbb{R}^{Nd \times 1}$, the global residual $\mathbf{R} \in \mathbb{R}^{Nd \times 1}$, and the global signal $\mathbf{S} \in \mathbb{R}^{Md \times 1}$ (where $M = |\mathcal{E}|$). While $\mathbf{U}$ represents the spatial distribution of the intensive density (e.g., concentration or momentum), $\mathbf{R}$ quantifies the instantaneous flux imbalance. By treating the residual as a thermodynamic force driving a continuous process of symmetry restoration, the HPN models the system's trajectory as the simultaneous co-evolution of the state $\mathbf{U}$ and the underlying transport geometry $\mathbf{W}$.

\subsection{Fundamental Invariants and Physical Consistency}
\label{sec:consistency}
We identify the fundamental structural properties that govern the evolution of a Hebbian Physics Network (HPN). These properties distinguish between invariants enforced exactly by the graph architecture and thermodynamic consistency conditions that characterize the class of physical systems the framework is designed to model.
\begin{enumerate}[label=\textbf{C\arabic*.}]
\item \textbf{Global Conservation as an Exact Structural Invariant.} The HPN architecture is constructed such that, for any configuration of the transport operator 
$\mathbf{W}$ and state $\mathbf{U}$, the sum of residuals vanishes identically:
\begin{equation}
    \sum_i R_i = 0.
\end{equation}
Because transport occurs exclusively through internal redistribution along directed edges, the total intensive state $\sum_i U_i$ is preserved algebraically. Global conservation is therefore an exact structural invariant of the network, rather than a property enforced asymptotically or approximated through iteration.

\item \textbf{Coercivity Condition and Symmetry Restoration.} For a physical system to admit a stable attractor, the mapping between thermodynamic forces (the residuals) and the state must satisfy a monotonicity condition consistent with positive entropy production. We characterize this requirement through a local coercivity condition on the residual operator:
\begin{align}
(\mathbf{R}(\mathbf{U}_a)-\mathbf{R}(\mathbf{U}_b))^\top& (\mathbf{U}_a-\mathbf{U}_b) \nonumber \\ &\ge c\|\mathbf{U}_a-\mathbf{U}_b\|^2_\mathbf{G},
\label{eq:local_coercivity}
\end{align}
for some $c > 0$, where $\mathbf{G}$ is a the nodal metric.
In this sense, the weights $\mathbf{W}$ function as a dynamic, metric-like structure that locally restores the dissipative symmetry required by the Second Law.
\end{enumerate}
\textbf{C2} defines a condition of thermodynamic admissibility. The class of systems considered in this work is restricted to those whose residual mapping satisfies this condition; systems that violate \textbf{C2} fall outside the scope of the HPN framework and cannot be consistently modeled within it. Accordingly, \textbf{C2} is treated as an axiom throughout. Sections~\ref{sec:scalar-onsager} -- \ref{sec:stability} show that bounded promotion, positive decay, and adiabatic separation ensure that the adaptive geometry realizes a locally monotone residual operator compatible with this thermodynamic admissibility.

\subsection{Constitutive Relaxation and the Onsager Limit}
\label{sec:scalar-onsager}
The HPN manifests physical dynamics through the simultaneous relaxation of the nodal state $\mathbf{U}$ and the transport geometry $\mathbf{W}$. This co-evolution is governed by two distinct internal timescales that characterize the substrate's response to thermodynamic stress.

\subsubsection{State Relaxation and the Kinetic Limit}
The nodal state $U_i$ evolves to dissipate local accumulations at a rate dictated by the substrate's information propagation speed. Utilizing the nodal metric $\Omega_i$ defined in Sec.~\ref{sec:problem}, the update is given by:
\begin{equation}
\Delta U_i = \frac{\eta_{u}}{\Omega_i} \left( \sum_{j \in \mathcal{N}(i)} (J_{ij} - J_{ji})\right).
\label{eq:state-update}
\end{equation}
Here, $\eta_u$ represents the kinetic relaxation rate. To ensure numerical stability and physical causality, $\eta_u$ is strictly bounded by the Courant–Friedrichs–Lewy (CFL) limit, ensuring that transport does not exceed the characteristic grid velocity of the discrete manifold.

\subsubsection{Structural Adaptation of the Transport Geometry}
The local transport geometry $W_{ij}$ evolves through a superposition of three mechanisms that reshape the pathway conductance in response to the residual $R_i$:
\begin{align}
\Delta W_{ij}= &-\underbrace{\eta_w f(R_i) S_{ij}}_{\text{Plastic Yield}} + \underbrace{\varepsilon g(|R_i|) S_{ij}}_{\text{Promotion}} - \underbrace{\lambda W_{ij}}_{\text{Decay}}.
\label{eq:w-update-general}
\end{align}
\begin{itemize}
\item The \textbf{Plastic Yield} term drives the structural adaptation of the substrate. If a specific pathway correlates with local accumulation, this term reduces its effective geometric conductance; conversely, pathways that aid in relieving stress are reinforced.
\item The \textbf{Promotion} term introduces a small baseline permeability that prevents premature elimination of transport pathways and maintains numerical and structural regularity of the operator near equilibrium. This term is auxiliary to dissipation and does not alter the direction of relaxation when residuals are significant.
\item The \textbf{Decay} term imposes a physical bound on the coupling strength, preventing the unbounded growth of transport coefficients and ensuring the weights remain within a physically admissible range.
\end{itemize}

\subsubsection{Gradient Descent Interpretation and Trajectory Invariance}
\label{sec:grad-descent}
To formalize the stability of the HPN, we define the global residual energy $\mathcal{L}$ as the squared norm of the residuals under the inverse nodal metric:
\begin{equation}
\mathcal{L} = \frac{1}{2} \mathbf{R}^\top \mathbf{G}^{-1} \mathbf{R} = \frac{1}{2} \sum_{i} \frac{1}{\Omega_i} R_i^2.
\label{eq:energy_functional}
\end{equation}
This functional represents the total `unresolved stress' within the substrate. The gradient of this energy with respect to a specific local weight $W_{ij}$ is given by:
\begin{equation}
\frac{\partial \mathcal{L}}{\partial W_{ij}} = \frac{R_i}{\Omega_i} \cdot \frac{\partial R_i}{\partial W_{ij}} = \frac{R_i}{\Omega_i} S_{ij}.
\label{eq:gradient_derivation}
\end{equation}
Comparing this result to the Plastic Yield term in the weight update rule [Eq.~(\ref{eq:canonical-weight-update})], we find that the anti-Hebbian update constitutes local gradient descent on the residual energy landscape. This interpretation guarantees that, irrespective of the instantaneous state evolution, the constitutive adaptation driven by plastic yield acts to reduce local thermodynamic stress.

A key consequence of this gradient structure is that the direction of the structural adaptation is determined solely by the local residual density and signal. While the mobility rate $\eta_w$ dictates the `computational velocity' along the path, it does not alter the path itself. 

\begin{lem}[Invariance of the Relaxation Trajectory]
\label{lem:lemma1}
Assume smoothness of the residual energy $\mathcal{L}$, bounded promotion and decay, and adiabatic separation between state relaxation and structural adaptation. Then, for any structural mobility $\eta_w$ within the stable bandwidth relative to the kinetic limit $\eta_u$, the HPN traverses a unique, physically defined trajectory in the $(\mathbf{U}, \mathbf{W})$ configuration space. The iteration index $n$ functions as an internal relaxation time, with $\eta_w$ acting solely as a reparameterization of progress along this trajectory.
\end{lem}
\begin{proof}
To prove that the physical trajectory of the substrate is independent of the structural mobility $\eta_w$, we express the full weight update [Eq. \eqref{eq:canonical-weight-update}] as a dynamic vector field. We define the Structural Mobility $\eta_w$ as the common pre-factor for the substrate's adaptive response, such that the promotion ($\varepsilon$) and decay ($\lambda$) are scaled relative to this mobility: $\varepsilon = \eta_w \hat{\varepsilon}$ and $\lambda = \eta_w \hat{\lambda}$.The continuous-limit evolution of the weights $\mathbf{W}$ with respect to the iteration index $n$ is:
\begin{equation}
\frac{d\mathbf{W}}{dn} = \eta_w \left[ -\nabla_{\mathbf{W}} \mathcal{L} + \hat{\varepsilon} \mathbf{\Gamma}_R(\mathbf{R}) - \hat{\lambda} \mathbf{W} \right]
\end{equation}
where $\nabla_{\mathbf{W}} \mathcal{L}$ is the gradient of the residual energy and $\mathbf{\Gamma}_R(\mathbf{R})$ represents the global promotion field. The unit tangent vector $\mathbf{\hat{v}}$ defining the direction of the trajectory in the manifold is:
\begin{align}
\mathbf{\hat{v}} = \frac{d\mathbf{W} / dn}{\|d\mathbf{W} / dn\|} &= \frac{\eta_w \left[ -\nabla_{\mathbf{W}} \mathcal{L} + \hat{\varepsilon} \mathbf{\Gamma}_R - \hat{\lambda} \mathbf{W} \right]}{\|\eta_w \left[ -\nabla_{\mathbf{W}} \mathcal{L} + \hat{\varepsilon} \mathbf{\Gamma}_R - \hat{\lambda} \mathbf{W} \right]\|} \nonumber \\ &= \frac{-\nabla_{\mathbf{W}} \mathcal{L} + \hat{\varepsilon} \mathbf{\Gamma}_R - \hat{\lambda} \mathbf{W}}{\|-\nabla_{\mathbf{W}} \mathcal{L} + \hat{\varepsilon} \mathbf{\Gamma}_R - \hat{\lambda} \mathbf{W}\|}
\end{align}
Since the scalar $\eta_w$ cancels out, the direction of the update at any point in phase space is determined exclusively by the relative balance between the physical gradient, the structural promotion, and the homeostatic decay.

We introduce a dimensionless path parameter $\tau$ to represent  the reparameterized progress of the substrate adaptation, such that $d\tau \propto \eta_w dn$, which implies: 
\begin{equation}
\frac{d\mathbf{W}}{d\tau} = -\nabla_{\mathbf{W}} \mathcal{L} + \hat{\varepsilon} \mathbf{\Gamma}_R - \hat{\lambda} \mathbf{W}
\end{equation}
This equation is entirely independent of $\eta_w$. It demonstrates that $\eta_w$ serves only as a reparameterization of the path length. Within the stable, adiabatic regime described above, the HPN traverses a unique relaxation trajectory that is invariant under rescaling of $\eta_w$. Within this regime, the emergent physics reflects a structural property of the system's energy landscape and its relative relaxation constants ($\hat{\varepsilon}, \hat{\lambda}$) rather than an artifact of the computational update rate. They physically characterize the material response of the simulated substrate, defining the balance between exploratory noise and structural stability.
\end{proof}

For the HPN to traverse a unique physical path independent of $\eta_w$, the structural mobility must satisfy the following bounds:
\begin{enumerate}
    \item The Upper Bound: The Adiabatic Stability Limit. The structural mobility $\eta_w$ is upper-bounded by the requirement that the geometry does not outpace the nodal relaxation. Mathematically, this is constrained by the curvature of the residual energy landscape $\mathcal{L}$. To maintain stability, the ``step size" in weight-space must not exceed the local characteristic scale of the gradient:\begin{equation}\eta_w < \frac{2}{\
    \Lambda_{\text{max}}(\mathbf{H}_W)}\end{equation}where $\Lambda_{\text{max}}(\mathbf{H}_W)$ is the largest eigenvalue of the Hessian of the residual energy with respect to the weights. Beyond this bound, the system enters an unstable regime where the updates $\Delta W$ overshoot the equilibrium, leading to unphysical oscillations or divergence. Physically, this represents the Structural Yield Limit of the substrate---if the material is forced to adapt faster than its internal state can settle, the physical continuity of the trajectory is broken.

    \item The Lower Bound: The Resolution Limit. Conversely, $\eta_w$ is lower-bounded by the practical requirement of numerical resolution. While the path remains theoretically invariant as $\eta_w \to 0$, an excessively small mobility leads to a ``stable" substrate where structural adaptation is negligible within the simulation's timeframe. For a unique path to be observable, $\eta_w$ must be large enough to allow the weights to escape the ``noise floor" created by the promotion term $\varepsilon$:
    \begin{equation}
    \eta_w > \frac{\Omega \varepsilon e^{-|\frac{R}{\Omega}|}}{|R|}
    \end{equation}
    This ensures that the Plastic Yield (driven by the residual) remains the dominant force over the Structural Promotion (driven by sub-scale fluctuations).
\end{enumerate}

\subsubsection{Canonical Choice and the Linear Limit}
\label{sec:canonical}
While the HPN framework admits any odd, monotone modulating function $f(\cdot)$ and non-negative, bounded function $g(\cdot)$, we adopt the linear response $f(R_i) = R_i/\Omega_i$ and the exponential gate $g(|R_i|) = e^{-|R_i/\Omega_i|}$ for the remainder of this work. This yields the canonical weight update:
\begin{equation}
\Delta W_{ij} = \underbrace{-\frac{\eta_w}{\Omega_i} R_i S_{ij}}_{\substack{\text{Plastic} \\ \text{Yield}}} ~~+~~ \underbrace{\varepsilon e^{-|\frac{R_i}{\Omega_i}|} S_{ij}}_{\substack{\text{Structural} \\ \text{Promotion}}}~~~ - \underbrace{\lambda W_{ij}}_{\substack{\text{Homeostatic } \\\text{Decay}}}.
\label{eq:canonical-weight-update}
\end{equation}
The linear choice $f(R_i) = R_i/\Omega_i$ corresponds to local gradient descent on the residual energy landscape, $\mathcal{L}$, while the exponential gate $g(|R|)$ modulates auxiliary structural activity on the magnitude of thermodynamic stress. To ensure physical consistency and numerical stability, we impose the hierarchical constraint $\varepsilon \ll \lambda$ such that promotion acts as a small perturbation relative to homeostatic decay.

Under this parameterization, the promotion term prevents structural degeneration of the transport geometry near equilibrium. If the promotion term is set to zero ($\varepsilon$ = 0), the weight update reduces to the plastic-yield and decay contributions. In regimes where residuals become small, the plastic term vanishes, leaving only the decay term, so that $\Delta W_{ij} = -\lambda W_{ij}$. In this case, all couplings decay exponentially toward zero, causing the learned operator to collapse to a degenerate, low-conductance state. Such collapse suppresses transport capacity and slows subsequent adaptation under renewed forcing. The promotion term maintains a small baseline permeability when residuals are weak, preventing this decay-driven collapse while remaining subordinate to dissipative adaptation when residuals are large ($\varepsilon \ll \lambda$). The exponential gate ensures a smooth transition between exploration near balance and strictly dissipative restructuring under strong imbalance, without introducing additional thresholds or discontinuities.

This canonical choice thus separates exploratory structural variability from dissipative relaxation in a controlled manner. In the following subsection, we show that in the near-equilibrium limit---where residuals are small and structural adaptation is quasi-static---this formulation admits a linear-response description consistent with classical non-equilibrium thermodynamics.

\subsubsection{The Onsager Limit}
\label{sec:onsager}
To formally ground the HPN framework within classical non-equilibrium thermodynamics, we demonstrate that the adaptive substrate recovers linear, reciprocal transport laws in the near-equilibrium limit. In this regime, the discrete operator does not merely converge numerically, but autonomously organizes into a configuration consistent with Onsager’s reciprocal relations.

We consider the near-equilibrium regime in which the transport geometry is quasi-static such that $\Delta W_{ij} \approx 0$. Given sufficiently small residuals ($|R_i| \ll 1$), we apply the canonical modulators established in Sec.~\ref{sec:canonical}: the linear response $f(R) = R/\Omega$ and a baseline promotion $g(|R|) \approx 1$. Under these assumptions, the stationary weight $W_{ij}^*$ is determined by the balance of structural yield, promotion, and homeostatic decay:
\begin{equation}
0 = -\eta_w \frac{R_i}{\Omega_i} S_{ij} + \varepsilon S_{ij} - \lambda W_{ij}^*
\end{equation}
In the near-equilibrium regime we further assume quasi-static adaptation, such that residuals vary slowly compared to structural relaxation and may be treated as locally fixed when solving this condition. Solving for the stationary weight yields
\begin{equation}
W_{ij}^* = \frac{1}{\lambda} \left[ \varepsilon S_{ij} - \eta_w \frac{R_i}{\Omega_i} S_{ij} \right]
\end{equation}
Substituting this stationary weight into the flux definition, $J_{ij} = W_{ij}^* S_{ij}$, provides the effective constitutive relation for the pathway:
\begin{equation}
J_{ij} = \underbrace{\frac{\varepsilon}{\lambda} S_{ij}^2}_{J_\text{base}} - \left( \frac{\eta_w}{\lambda} S_{ij}^2 \right)\frac{R_i}{\Omega_i}
\label{eq:equi_flux}
\end{equation}
Here, $J_\text{base}$ represents a small baseline flux arising from the structural promotion. Since the canonical hierarchy enforces $\varepsilon \ll \lambda$ as defined in Section \ref{sec:canonical}, this term becomes negligible relative to the dissipative response. Consequently, the constitutive relation simplifies to a pure linear response:
\begin{equation}
J_{ij} \approx -L_{ij}R_i ,
\qquad
L_{ij}=\frac{\eta_w}{\lambda \Omega_i}S_{ij}^2 .
\end{equation}
The coefficient $L_{ij}$ thus plays a role of an effective local Onsager transport coefficient. In the near-equilibrium regime, neighboring nodes sample the same local thermodynamic state. Since, $S_{ij}=U_j$ is an intensive density and $\Omega_i$ is a slowly varying local capacity, we have $U_i = U_j + O(\nabla U)$ and $\Omega_i \approx \Omega_j + O(\nabla\Omega)$. Consequently, 
\[S_{ij}^2/\Omega_i = S_{ji}^2/\Omega_j + O(\nabla U, \nabla\Omega) \]
so the effective transport coefficients become symmetric to leading order:
\begin{equation}
L_{ij} \approx L_{ji}.
\end{equation}
Here, reciprocity refers specifically to edge-pair equality of the dissipative force–flux coupling between conjugate transport pathways. Any edge-pair inequality   in the learned operator corresponds to circulatory transport, which is energetically neutral and does not contribute to entropy production, fully consistent with Onsager’s original formulation. The recovery of this symmetry demonstrates that HPN adaptation is governed by a principle of functional reversibility: without explicit global enforcement, the adaptive substrate self-organizes reciprocal dissipative pathways in the near-equilibrium limit. Classical linear transport laws thus emerge naturally as the limiting description of residual-driven constitutive relaxation.

\subsection{Global Stability and Structural Exploration }
\label{sec:stability}
To establish the dynamical integrity of the HPN, we examine the evolution of the global residual energy, $\mathcal{L} = \frac{1}{2}\mathbf{R}^\top \mathbf{G}^{-1} \mathbf{R}$, which quantifies unresolved thermodynamic stress. While classical linear systems often require a monotonic Lyapunov decay, the HPN admits a more complex, non-monotonic energy landscape. Such behavior is naturally permitted --- and, in many non-linear systems necessary --- for structural exploration; it enables the substrate to navigate non-linear phase transitions where the transient accumulation of stress is a precursor to the discovery of dissipative pathways.

\subsubsection{The Dynamic Landscape and Co-evolution}
The total variation of the residual energy is governed by the coupled evolution of the nodal states and the transport geometry:
\begin{equation}
\frac{d\mathcal{L}}{dn} = \underbrace{\left( \nabla_{\mathbf{U}} \mathcal{L} \right)^\top \frac{d\mathbf{U}}{dn}}_{\text{Kinetic Dissipation}} + \underbrace{\left( \nabla_{\mathbf{W}} \mathcal{L} \right)^\top \frac{d\mathbf{W}}{dn}}_{\text{Structural Adaptation}}
\end{equation}
The first term represents the kinetic (state) dissipation. Substituting the state update
\[
\Delta \mathbf{U} = -\eta_u \mathbf{G}^{-1} \mathbf{R},
\]
and assuming that the coercivity condition (\textbf{C2}) holds locally for the current transport geometry, we obtain
\begin{align}
\left( \nabla_{\mathbf{U}} \mathcal{L} \right)^\top \frac{d\mathbf{U}}{dn}
&= \mathbf{R}^\top \mathbf{G}^{-1}
\left( \frac{\partial \mathbf{R}}{\partial \mathbf{U}} \right)
\left( -\eta_u \mathbf{G}^{-1} \mathbf{R} \right) \nonumber \\
&= -\eta_u \mathbf{R}^\top \mathbf{M}_{\mathrm{sym}} \mathbf{R} \nonumber \\
&\le 0,
\end{align}
where
\[
\mathbf{M}_{\mathrm{sym}}
\equiv
\mathbf{G}^{-1}
\left(
\frac{1}{2}
\left[
\frac{\partial \mathbf{R}}{\partial \mathbf{U}}
+
\left(
\frac{\partial \mathbf{R}}{\partial \mathbf{U}}
\right)^\top
\right]
\right)
\mathbf{G}^{-1}
\]
denotes the symmetric part of the weighted Jacobian. 
Under the coercivity/monotonicity condition (\textbf{C2}), 
$\mathbf{M}_{\mathrm{sym}}$ is positive semidefinite, implying
\[
\mathbf{R}^\top \mathbf{M}_{\mathrm{sym}} \mathbf{R} \ge 0.
\]
Consequently, for fixed transport geometry $\mathbf{W}$, 
state relaxation is dissipative with respect to the residual energy $\mathcal{L}$.

The second term introduces non-monotonicity. Because $\mathbf{R}$ is a non-linear function of both $\mathbf{U}$ and $\mathbf{W}$, the landscape itself shifts as the weights evolve. In regions of strong non-linearity, such as the onset of turbulence or vortex formation, the structural adaptation may temporarily increase $\mathcal{L}$. Physically, this represents the system climbing a structural barrier in the configuration space, accumulating stress in order to break an inefficient symmetry and access new transport pathways. 

For vector fields ($d > 1$), this co-evolution allows $\mathbf{W}$ to decompose into symmetric and skew-symmetric components. While the symmetric part drives the dissipation of residual energy, the skew-symmetric part organizes rotational transport allowing the system to redistribute flux and satisfy local conservation without additional energetic cost to the residual landscape. This mechanism is demonstrated in the lid-driven cavity results presented in Sec.~\ref{sec:cavity}. where purely viscous dissipation becomes insufficient and convective pathways emerge through a structural bifurcation.

\subsubsection{Global Boundedness and Lagrange Stability}
Although the HPN does not enforce global Lyapunov monotonicity, the structural dynamics remain bounded when $\lambda > 0$ and the promotion term is bounded. 
The structural update may be written as
\begin{equation}
\frac{d\mathbf{W}}{dn}
=
-\eta_w \nabla_{\mathbf{W}} \mathcal{L}
-\lambda \mathbf{W}
+\varepsilon e^{-|\mathbf{R}|}\mathbf{S}.
\label{eq:W_update}
\end{equation}

To assess boundedness, we consider the evolution of the weight norm. 
Using the identity
\[
\frac{d}{dn}\|\mathbf{W}\|
=
\frac{\mathbf{W}^\top}{\|\mathbf{W}\|}
\frac{d\mathbf{W}}{dn},
\]
and applying the Cauchy--Schwarz inequality together with 
$e^{-|\mathbf{R}|} \le 1$, we obtain
\begin{align}
\frac{d}{dn}\|\mathbf{W}\|
&\le
-\lambda \|\mathbf{W}\|
+\eta_w \|\nabla_{\mathbf{W}} \mathcal{L}\|
+\varepsilon \|\mathbf{S}\|.
\label{eq:W_norm_bound}
\end{align}

In the regimes considered here, $\|\nabla_{\mathbf{W}} \mathcal{L}\|$ remains bounded whenever the state and residual remain bounded, and $\|\mathbf{S}\|$ is bounded since it is constructed from local state variables. 
Hence there exists a constant $C>0$ such that
\begin{equation}
\frac{d}{dn}\|\mathbf{W}\|
\le
-\lambda \|\mathbf{W}\| + C.
\end{equation}

By standard comparison arguments for first-order differential inequalities, this implies
\begin{equation}
\limsup_{n\to\infty}\|\mathbf{W}_n\|
\le
\frac{C}{\lambda},
\end{equation}
and therefore the trajectory remains confined to a compact region of weight space. 
The coupled $(\mathbf U,\mathbf W)$ dynamics are thus Lagrange stable under bounded promotion and positive decay.

\section{Demonstrative Cases}
To illustrate the scope of Hebbian Physics Networks (HPNs), we present two representative systems: continuum diffusion and incompressible flow in a lid–driven cavity. In both cases, physically consistent structures emerge from random or perturbed initial states through local residual relaxation and anti-Hebbian weight adaptation---without explicitly solving the governing equations using the time as a global time marching problem.

Throughout the remainder of this work, variables and parameters are reported in dimensionless form to facilitate scale-independent analysis. Unless explicitly stated otherwise, all quantities have been normalized by their respective characteristic values as defined by the underlying conservation laws. Additionally, all simulations were conducted with $\eta_w$ chosen within the adiabatic stability bandwidth described in Section~\ref{sec:scalar-onsager}.
\subsection{Diffusion}
\label{sec:diffusion}

We consider an unstructured graph whose connectivity is defined by a local neighborhood criterion: each node \(i\) is connected to a node \(j\) if the Chebyshev distance between their coordinates satisfies
\[
\lVert \mathbf{x}_i - \mathbf{x}_j \rVert_{\infty} \le k,
\]
(see Sec.~\ref{sec:problem}). On this graph, we consider the capacity of all the nodes is equal such that $\Omega_i \equiv \Omega = 1$. The nodal state \(U_i\) is identified with the concentration \(c_i\), consistent with the definition of state in Sec.~\ref{sec:state_def} as the density of a conserved quantity, i.e. amount of substance in this case. The constitutive signal is taken to be the source concentration at the neighboring node, such that
\begin{equation}
    S_{ij} = c_j~ ,
\end{equation}
as the state-centered signal defined in Sec.~\ref{sec:state_def}. 

All constitutive physics is encoded in the adaptive transport operator $W_{ij}$. Therefore, the directed flux along each edge is defined as
\begin{align}      
    J_{ij} &= D W_{ij} S_{ij}.     
\end{align}
Here, the diffusivity $D$ is treated as a material property that sets the overall timescale and magnitude of diffusive transport, while the adaptive operator $W_{ij}$ encodes the effective transport geometry of the graph. Although $W_{ij}$ adapts in response to residuals, it remains dimensionless and represents relative pathway strengths rather than absolute transport rates. Keeping $D$ explicit therefore separates material properties from geometric structure, preventing degeneracy in the
learned operator and allowing the same transport geometry to be applied across different diffusive regimes. Absorbing $D$ into $W_{ij}$ would be mathematically equivalent but would obscure the
distinction between learned geometry and intrinsic material response.
The continuity residual at node $i$ is given by the discrete divergence
\begin{align}
    R_{c,i} &= D\sum_{j \in \mathcal{N}(i)} (W_{ij}S_{ij} - W_{ji}S_{ji}),     
\end{align}
which represents the local accumulation of concentration due to imbalance between incoming and outgoing transport pathways. 

In this case, transport between neighboring nodes $i$ and $j$ is represented by two directed couplings, $W_{ij}$ and $W_{ji}$. We impose an \emph{edge-pair equality} constraint,
\begin{equation}
    W_{ij} = W_{ji},
\end{equation}
which restricts the admissible transport geometry to link-level balanced couplings. This is a constitutive choice for the diffusion example: edge-pair unequal couplings generate divergence-free circulation and do not contribute to accumulation of a scalar conserved density. By enforcing edge-pair equality, all transport pathways contribute to local accumulation and relaxation. Under this restriction, the nodal residual reduces to a graph Laplacian acting on the concentration field, and diffusive transport emerges without explicitly encoding gradients in the signal.

Classical graph-based discretizations prescribe $W_{ij}$ directly from the geometric embedding (typically $ W_{ij} \propto \mathbf{r}_{ij}/\ |\mathbf{r}_{ij}\|^{2} $) and treat it as fixed in time.  Here, no geometric information is assumed \emph{a priori}: the effective geometry is inferred through the learned weights $W_{ij}$ themselves.

The node and weight updates follow the local anti-Hebbian rule
\begin{align}
    c_i^{n+1} &\leftarrow c_i^{n} + \eta_c\, R_{c,i}^{n}, \\
    W_{ij}^{n+1} &\leftarrow W_{ij}^{n}
      + S_{ij}\left(-\eta_w R_{c,i}^{n}
      + \varepsilon_w e^{-|R_{c,i}^{n}|}\right)
      - \lambda\, W_{ij}^{n},
\end{align}
where $S_{ij}$ is the \textit{local signal}, $R_{c,i}$ acts as a modulatory signal, and the decay term ensures long-term boundedness of the couplings. Superscripts, $(\cdot)^n$, denote the iteration index. 

The parameters used here are
\begin{align*}
&\eta_w=0.01,\quad
\varepsilon_w=0.0001, \quad
\eta_c^0=0.001,\quad \\
&\eta_c = \min \left( \frac{C|r_{ij}| |S_{max}|}{|J_{max}|},  \eta_{c}^0 \right), \quad
C = 1.0, \quad
\lambda=0.01,
\end{align*}
where, ${C|r_{ij}| |S_{max}|}/{|J_{max}|}$ is a CFL-like limit for diffusion transport.
The concentration field is initialized to zero, and a localized perturbation is introduced at $n=0$; all weights begin at unity.

\begin{figure*}[p] 
    \centering
    \includegraphics[width=\textwidth]{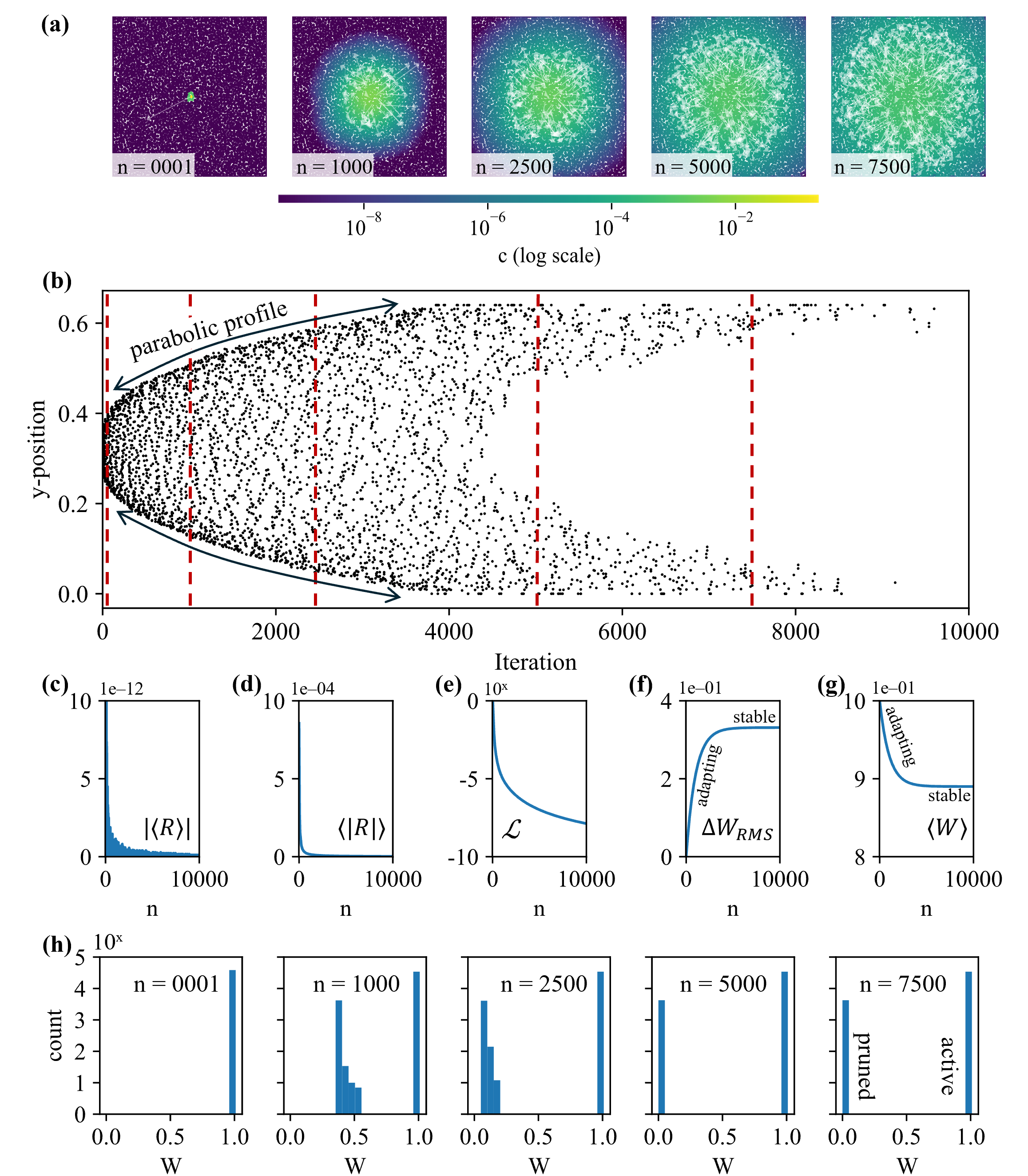}    
    \caption{\textbf{Residual-driven emergence of diffusive transport in a HPN.} (a) Evolution of the concentration field on an unstructured graph. (b) Spatiotemporal trajectory of the emergent transport front obtained by tracking the $y$-coordinates of nodes crossing a threshold concentration; red dashed lines indicate the iteration indices corresponding to the snapshots in (a). (c) Global conservation error $|\sum_i R_{c,i}|$, which remains at $\mathcal{O}(10^{-11})$ throughout the evolution, demonstrating exact global conservation. (d) Mean absolute local residual $\langle |R_c| \rangle$, which remains finite($\sim\mathcal{O}(10^{-5})$) and reflects persistent local thermodynamic stress. (e) Lyapunov function $\mathcal{L}$ showing exponential monotonic decrease with iterations. (f) Root-mean-square change in weights, $\Delta W_{\mathrm{RMS}}$, indicating rapid early adaptation followed by asymptotic stabilization of the constitutive operator. (g) Mean weight magnitude $\langle W \rangle$ approaching a stable value as the learned transport geometry statistically becomes stationary. (h) Histograms of the weight distribution $W_{ij}$ at selected iterations, illustrating pruning of inactive pathways and emergence of an effective transport geometry.}
    \label{fig:diffusion_hpn}
\end{figure*}

Figure~\ref{fig:diffusion_hpn} summarizes the resulting dynamics.  Panel (a) shows the evolution of the concentration field.  The spreading front co-evolves with the operator: the fluxes (white quivers) are generated by $W$, and those fluxes in turn reshape $W$ through the anti-Hebbian update.  This bidirectional coupling produces a finite-speed propagation of activity.  The structure of the diffusion front in panel~(b)---obtained by tracking the $y$-coordinates of first activations---follows a clear parabolic envelope, $r\propto\sqrt{n}$. This observation allows the iteration index 
$n$ to be calibrated against physical time \textit{a posteriori}, rather than being imposed externally.

The global conservation error in panel~(c) remains at machine precision throughout the evolution, demonstrating that conservation is invariant and not an emergent property. Relaxation instead occurs locally: the mean absolute residual in panel~(d) shows an initial burst followed by rapid decay as the transport geometry adapts to redistribute local imbalances. 

The Lyapunov function $\mathcal{L} = \tfrac{1}{2}\sum_i \|R_{c,i}\|^2$ in panel~(e) decreases monotonically and then approaches a small, finite plateau. 
For unforced diffusion on a closed (periodic) domain, the physical equilibrium corresponds to a spatially uniform concentration and vanishing macroscopic flux. 
In the HPN implementation, however, the bounded promotion term and finite numerical resolution maintain a residual noise floor, so the nodal residuals need not vanish identically even when the concentration field is effectively uniform. 

The long-time behavior therefore represents a stationary fluctuation regime around the equilibrium manifold rather than a thermodynamic non-equilibrium steady state. 
Global conservation is satisfied exactly (panel~(c)), and the learned transport geometry $W$ stabilizes in norm, while the small but non-zero $\Delta W$ reflects continued microscopic reorganization driven by bounded promotion rather than sustained macroscopic transport. The finite plateau of $\mathcal{L}$ thus reflects the noise-sustained residual level inherent to the adaptive dynamics, not the presence of persistent physical flux.

Panels~(f) and (g) show the root-mean-square change in weights and the mean weight magnitude, respectively: $W$ adapts substantially over the first $\sim\!4000$ iterations and then asymptotically stabilizes, marking the emergence of a steady transport geometry.  The histograms in panel~(h) reveal the pruning behavior characteristic of the weight update rule: weights that do not contribute to residual reduction shrink toward zero, producing a bistable distribution that encodes the effective transport geometry of the learned medium.

Although the algebraic form of the update resembles a graph discretization of diffusion, the mechanism of evolution is fundamentally distinct from explicitly time-integrating the parabolic PDE $\partial_t c = D\nabla^2c$ under an external clock. No global timestep is imposed, and no fixed Laplacian is ever constructed or inverted. Instead, both state $c$ and transport geometry $W$ evolve together: $W$ determines the instantaneous fluxes; the fluxes produce the residual; and the residual drives the joint update of the state and the operator.  Each iteration is therefore not a step of a numerical time-marching solver for $\partial_t c$, but a local relaxation event in which state and transport geometry co-evolve.  As $W$ stabilizes, the emergent dynamics reproduce the diffusion law as a coarse-grained description, but the computation itself is the relaxation trajectory realized through adaptive local transport.

An explicit comparison of this method with the classical diffusion solution, carried out for a minimal two-node system, is provided in Sec. S1 A of the Supplemental Information \cite{supp_info}.

\subsection{Incompressible flow: lid–driven cavity}
\label{sec:cavity}
\begin{figure*}
    \centering
    \includegraphics[width=0.85\textwidth]{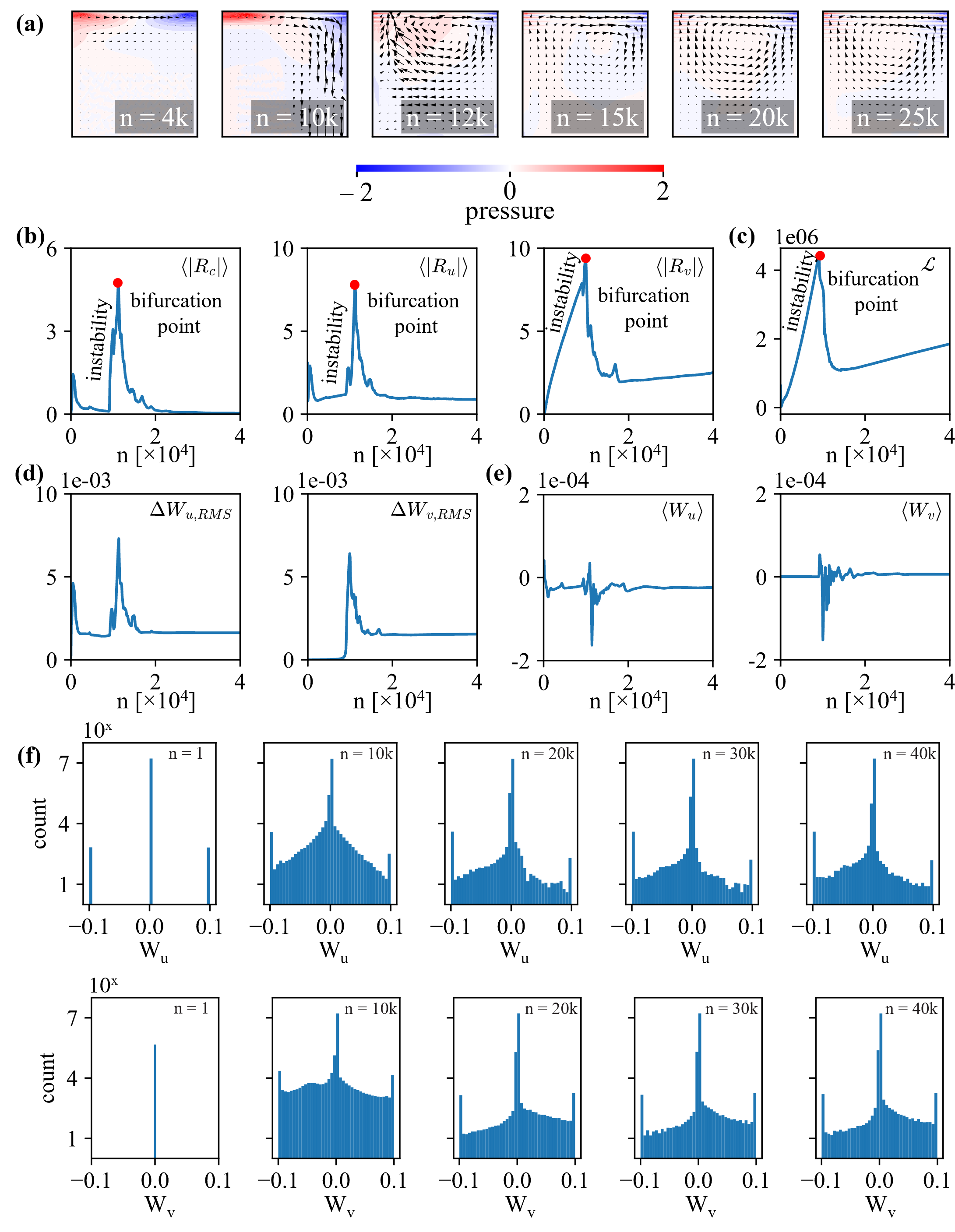}
    \caption{\textbf{Residual-driven flow formation in a HPN.} (a) Streamline snapshots for the lid–driven cavity. From a nearly quiescent state, the network organizes the velocity into the characteristic recirculating pattern; pressure is shown as a background colormap, centered at zero. (b) Mean continuity, streamwise, and cross-stream residuals $\langle|R_c|\rangle$, $\langle|R_u|\rangle$, and $\langle|R_v|\rangle$, showing an early surge followed by rapid decay as imbalances are redistributed through the learned transport operator. (c) Lyapunov function $\mathcal{L}=\tfrac12\sum_i (R_{c,i}^2 + R_{u,i}^2 + R_{v,i}^2)$, increases until the network discovers the vertical degree of freedom and bifurcates after reaching a maximum. (d) Root-mean-square weight changes $\Delta W_{u,\mathrm{RMS}}$ and $\Delta W_{v,\mathrm{RMS}}$, indicating an initial phase of strong plasticity followed by stabilization. (e) Mean weights $\langle W_u\rangle$ and $\langle W_v\rangle$ approaching stable values as the transport geometry settles.  (f) Histograms of $W^{(u)}_{ij}$ and $W^{(v)}_{ij}$ at selected iterations, initially narrow, then broadening during adaptation and developing heavy-tailed structure associated with shear, before the flow settles into a non-equilibrium steady state.
    }
    \label{fig:ldc_hpn}
\end{figure*}

In this example, the residual operator is defined by the discrete Navier--Stokes force balance. The objective is not to re-derive the fluid equations from transport incompatibility alone, but to examine whether the HPN can adapt its constitutive transport geometry to realize the correct nonlinear dynamics under physically meaningful constraints. 

Accordingly, we employ the HPN in a hybrid constraint--transport formulation. This separation is necessary because incompressibility and momentum balance are strict physical constraints that must be enforced with fidelity and do not arise in scalar diffusion problems. In this setting, the Navier--Stokes operator is used solely to evaluate the degree to which the current state violates force balance and divergence-free conditions. It does not prescribe the transport mechanism, nor is it integrated forward in time (i.e., no explicit $\partial_t \mathbf{u}$ time-marching is performed). Instead, the HPN adapts the constitutive transport geometry $W$ responsible for momentum redistribution, while the residual operator defines the physical constraints to be satisfied.

For incompressible flow, two conserved quantities are involved: mass and momentum. The fundamental state variables are therefore the mass density $\rho$ and the momentum density $\rho\mathbf{u}$. In the incompressible limit, $\rho$ is uniform and fixed, so all dynamical evolution occurs in the velocity field,
\begin{equation}
    U_i \equiv \mathbf{u}_i = (u_i, v_i),
\end{equation}
which fully characterizes the local momentum density up to a constant factor. Consistent with the state-centered formulation introduced earlier, the signal transmitted along a directed edge $(j \to i)$ is the velocity at the source node,
\begin{equation}
    S_{ij} \equiv \mathbf{u}_j .
\end{equation}
Momentum is redistributed through an adaptive transport operator $W_{ij}$ acting on this signal.

The governing equations are used solely to evaluate constraint residuals. In particular, the Navier--Stokes operator is employed to compute the local imbalance of momentum and the local violation of incompressibility. On an unstructured graph, spatial derivatives are approximated using discretized Green--Gauss operators constructed from the local neighborhood of each node. The resulting residuals are
\begin{align}
    \mathbf{R}_{u} &= \rho(\mathbf{u}\cdot\nabla)\mathbf{u}
    - \nabla p - \mu\nabla^2\mathbf{u}, \nonumber \\
    R_{c} &= \rho\nabla\cdot\mathbf{u},
\end{align}
where $p$ is the pressure and $\mu$ is the viscosity. Since the density $\rho$ is constant, it may be absorbed into the residual normalization; it is retained here only to emphasize momentum balance and is absorbed hereafter for notational clarity.

In this formulation, pressure is not treated as an independently transported quantity. Instead, it arises as a response to violations of mass conservation. Local divergence of the velocity field signals a breakdown of incompressibility, and the pressure field acts as a Lagrange multiplier that corrects this violation by redistributing momentum. Pressure therefore appears only through the continuity residual and is updated as part of the residual relaxation process.

The relaxation dynamics are governed by an adaptive vector transport operator $W_{ij}$. Because the nodal state is two-dimensional, the learned transport geometry at each edge is represented as a $2\times2$ operator acting on the velocity vector,
\begin{equation}
    W_{ij} =
    \begin{pmatrix}
        W_{x,ij} & 0 \\
        0 & W_{y,ij}
    \end{pmatrix},
\end{equation}
where $W_{x,ij}$ and $W_{y,ij}$ encode transport of the $x$- and $y$-velocity components, respectively. This block-diagonal form preserves independent transport of each component while allowing directional bias through edge-pair inequality ($W_{ij}\neq W_{ji}$). As in the diffusion case, the weights define a local interaction graph of finite radius around each node and collectively encode the learned transport geometry.

Although the residuals are computed from the Navier--Stokes constraints, the adaptation mechanism itself is unchanged: transport pathways that reduce local momentum imbalance are reinforced, while those that oppose relaxation are suppressed. The weights evolve according to a residual-modulated anti-Hebbian rule,
\begin{equation}
    W_{ij}^{n+1}
    = (1-\lambda) W_{ij}^{n}
    + \mathbf{u}_j^{n} \otimes
    \left(
        -\eta_w \mathbf{R}_{u,i}^{n}
        + \varepsilon_w g(-|\mathbf{R}_{u,i}^{n}|)
    \right),
\end{equation}
where the modulation function $g(\cdot)$ acts component-wise so that promotion occurs independently along each velocity component.

For incompressible flow, the primary constraint is mass conservation. Local violations of incompressibility are quantified by the continuity residual $R_{c,i}$. Rather than being applied directly to the velocity field, this residual is enforced through a pressure correction: pressure acts as a Lagrange multiplier that restores mass balance,
\begin{equation}
p_i^{n+1} = p_i^{n} + \eta_p R_{c,i}^{n}.
\end{equation}
This update defines the admissible manifold on which momentum may be redistributed.

Momentum relaxation is then realized through adaptive transport along this pressure-corrected manifold. The velocity field evolves according to
\begin{equation}
\mathbf{u}_i^{n+1}
= \mathbf{u}_i^{n}
+ \eta_u \sum_j W_{ij}^{n}\mathbf{u}_j^{n},
\end{equation}
which represents the incoming contribution of transported momentum from neighboring nodes. A fully conservative net-flux formulation would also include an outgoing contribution of the form $\eta_u \sum_j W_{ji}^{n}\mathbf{u}_i^{n}$, accounting for momentum transported away from node $i$. In the present hybrid approach, this term is not applied explicitly in the velocity update. Instead, its effect is captured indirectly: momentum outflow that induces local divergence is measured by the continuity residual $R_{c,i}$ and corrected through the pressure update.

In this way, mass conservation and momentum transport are handled by complementary mechanisms. Pressure corrections enforce incompressibility as a hard constraint, while adaptive transport redistributes momentum within that constraint. The velocity update is thus not a standalone conservative step, but part of a coupled relaxation in which pressure and transport geometry jointly determine the admissible dynamics.

Figure~\ref{fig:ldc_hpn} summarizes the relaxation dynamics of the HPN applied to the lid--driven cavity. Panel~(a) shows streamline snapshots at selected iterations. Starting from an initially quiescent field, the flow organizes into the canonical recirculating structure as the learned transport operator redistributes momentum along directions indicated by local residual imbalance. The background pressure field (centered at zero) reflects the progressive establishment of shear-driven pressure gradients, while the quiver arrows show the emergence and sharpening of the primary vortex.

The oscillatory pressure features that appear near the top corners during early iterations do not indicate numerical instability. Rather, they are transient signatures of enforcing incompressibility through residual-driven local adjustments without a global elliptic projection. The resulting pressure fluctuations are consistent with pseudo-compressible (artificial compressibility) relaxation mechanisms \cite{chorin1997numerical} and decay as the learned transport geometry stabilizes.

Panel~(b) tracks the evolution of the constraint residuals.
The continuity residual $\langle|R_c|\rangle$ and the momentum residual components $\langle|R_u|\rangle$ and $\langle|R_v|\rangle$ do not relax monotonically. Instead, they exhibit a pronounced transient surge followed by rapid decay. This surge marks a bifurcation in the relaxation dynamics of the coupled state–operator system $(U,W)$. As the network begins simultaneously enforcing incompressibility and momentum balance, the initial transport geometry becomes incompatible with the imposed constraints, leading to an accumulation of local stress. Once this stress exceeds a threshold, the system undergoes a bifurcation in which the initial relaxation pathway loses stability and a new dynamical regime---characterized by transverse momentum  redistribution---emerges. The Lyapunov function in panel~(c), $\mathcal{L} =  \tfrac12\sum_i\!\left( R_{c,i}^2+R_{u,i}^2+R_{v,i}^2 \right)$, exhibits the same structure. The sharp peak (red marker) identifies the bifurcation point at which the transport operator reorganizes to admit admissible pathways for residual relaxation. Following this bifurcation, residuals decay as the system converges toward a stable non-equilibrium steady state supported by the learned transport geometry.

Panels~(d) and (e) display the evolution of the learned transport operator. The root-mean-square weight changes $\Delta W_{u,\mathrm{RMS}}$ and $\Delta W_{v,\mathrm{RMS}}$ show an initial phase of strong plasticity followed by asymptotic freezing as the operator adapts to the dominant shear and recirculation pattern. The mean weights $\langle W_u\rangle$ and $\langle W_v\rangle$ likewise settle into steady values as the learned transport geometry stabilizes.

Panel~(f) shows histograms of the weight distributions $W^{(u)}_{ij}$ and $W^{(v)}_{ij}$ at selected iterations. The initially narrow distributions broaden during flow formation and develop heavy-tailed structure as transport pathways become selectively reinforced. As the steady vortex emerges, couplings that do not contribute to residual reduction, shrink toward zero, while pathways aligned with active momentum redistribution remain strengthened. This pruning and anisotropic reinforcement reflect the emergence of an effective constitutive transport geometry tailored to the cavity constraints.

A quantitative comparison of the converged steady state at $\mathrm{Re}=100$ with the standard benchmark data of Ghia et. al. \cite{ghia1982high} is provided in Sec. S2 of the Supplemental Information \cite{supp_info}, confirming agreement with the classical lid--driven cavity solution.

Overall, this example highlights that the object learned by the HPN is the local transport geometry, not the Navier--Stokes operator itself. Fixed discrete Navier--Stokes operators (used only for residual evaluation) supply the momentum and incompressibility constraints, while the evolving weights $W$ determine how momentum is redistributed from one relaxation event to the next. Although the resulting flow field reproduces the characteristic lid--driven cavity structure, the trajectory is not a time integration of the PDE; rather, the computation is the equilibration process itself, mediated by a plastic transport operator that continuously reshapes its geometry to relieve constraint-induced stress.

\section{Discussion and Outlook}
\label{sec:discussion}

The Hebbian Physics Network (HPN) shares the objective of physics-informed machine learning (PIML): to embed physically meaningful structure directly into computational systems so that their evolution respects conservation principles and constitutive constraints. The distinction lies in the mode of enforcement. In most PIML approaches, physical structure is imposed top-down, typically through fixed governing equations or global loss functions evaluated over a prescribed operator. In HPN, by contrast, physical structure is adapted bottom-up: the system reshapes its local constitutive pathways directly in response to observed imbalance, without requiring global constraint enforcement or a fixed constitutive operator.

An instructive analogy is a two-dimensional grassy field exposed to airflow. One approach to reconstructing the flow is to assume the Navier–Stokes equations in advance and fit a velocity field consistent with the observed bending of the grass \cite{raissi2020hidden}. This is a top-down strategy: the governing equation is fixed, and observations are used to determine the state. An alternative is to infer how each grass blade responds locally to the surrounding motion using the observed blade deflections as boundary information\cite{cardona2021wind}. No global flow equation is imposed as a synchronized evolution law. Instead, a local constitutive description is described everywhere, and the global flow field emerges as a consequence of these local responses.

HPN follows this second paradigm. It does not encode the Navier–Stokes equations as a fixed transport operator or time-integration rule; rather, it adapts the transport structure to be consistent with local observations, and the familiar continuum equations reappear only as an emergent, coarse-grained description of the resulting dynamics.

HPN provides a complementary formulation of transport dynamics, whose relaxed coarse-grained behavior is described by dynamic partial differential equation (PDE) solvers. In this view, PDEs provide macroscopic conservation laws, while dynamic PDE solvers enforce these laws through globally synchronized update rules. By requiring instantaneous consistency across the domain, many traditional PDE solvers necessarily coarse-grain the underlying non-equilibrium dynamics, masking the finite-rate local relaxation processes inherent in physical substrates \cite{lebon2008understanding, hoover2012computational, cross1993pattern}.

This bottom-up perspective places HPN closer in form to energy-based computation, in which system evolution is governed by internally generated consistency signals rather than externally prescribed update rules \cite{Hopfield1982, Ackley1985, LeCun2006}. The essential distinction is that, in HPN, these signals are physically interpretable as indicators of local non-equilibrium structure. In non-equilibrium thermodynamics, residuals---the local imbalances in a conserved quantity---signal that the available transport pathways are incompatible with the instantaneous state \cite{Onsager1931, Prigogine1977}. HPN treats these residuals as the drivers of constitutive adaptation: transport pathways reorganize until local inconsistencies are reduced. In HPN, this constitutive adaptation is realized explicitly through the evolution of the transport weights $W_{ij}$ that define how state is communicated between neighboring nodes.

Formally, we denote by $\mathcal{W} = \{W_{ij}\}$ the collection of transport weights, and by $\mathbf{W} \in \mathbb{R}^{N \times N}$ the sparse operator assembled from these weights. Rather than viewing $\mathbf{W}$ as a fixed discretization of a differential operator, it is more natural to interpret it as a discrete constitutive structure defined on the graph. In this interpretation, the weights $W_{ij}$ play a role structurally analogous to gauge connections: they specify how state information is transported between neighboring nodes and how local reference frames are coupled, in the same sense that connections define parallel transport on a manifold \cite{misner1973gravitation, arnold2012geometrical}. 

From this perspective, the residual at a node reflects the failure of these local connections to be mutually compatible. This is directly analogous, at a discrete level, to the appearance of curvature in geometric theories, where incompatibility of parallel transport around infinitesimal loops signals a non-flat manifold \cite{hehl1976general, nakahara2018geometry}. The residual thus serves as an operational indicator of geometric inconsistency in the learned transport structure, rather than as a numerical error in state variables.

From this structural standpoint, the dynamics of HPN can be interpreted as a process of local geometry adjustment. When residuals are large, the constitutive connections are poorly aligned, and the effective manifold defined by $\mathbf{W}$ is strongly curved. As the system evolves, updates to $\mathcal{W}$ act to flatten this local geometry, reducing curvature and improving compatibility between neighboring states. This adaptive reshaping of transport structure bears a close conceptual relationship to geometric formulations of dynamics and natural-gradient flows, where evolution proceeds along directions determined by the local metric structure \cite{amari1998natural, marsden2013introduction}.

The precise mathematical correspondence between residuals, curvature, and gauge structure is not pursued here and will be developed in future work. The present discussion is intended to provide a structural and physical interpretation of adaptive transport in HPN, rather than a formal gauge-theoretic construction.

Taken together, these observations position the Hebbian Physics Network as a transparent physics-informed learning framework in which transport dynamics are organized through local resolution of imbalance rather than enforced by a globally prescribed evolution law. In this view, dissipation, constitutive structure, and symmetry restoration emerge jointly from residual-driven adaptation, with the learned transport geometry encoding how a system redistributes stress as it relaxes. Classical partial differential equations then appear as macroscopic descriptions of the relaxation trajectories generated by the HPN, rather than as equations that are explicitly time-integrated within the algorithm. The HPN framework is therefore not intended to replace continuum theory, but to clarify how constitutive structure consistent with such descriptions can arise dynamically from local interactions --- an outlook that becomes particularly relevant far from equilibrium, where effective transport laws are not known in advance and must be discovered through the dynamics themselves.

\begin{acknowledgments}
The open access publication of this work was supported by JSPS KAKENHI Grant Number JP23K28154 and JST CREST Grant Number JPMJCR24R2.

G.A. gratefully acknowledges his partner, Ms. Shambhavi Chaturvedi (National Institute of Genetics, Japan), for insightful discussions, especially on the neuroscience aspect relevant to this work.
\end{acknowledgments}



\end{document}